\documentstyle[twocolumn,prl,aps]{revtex}
\begin{document}

{\bf Comment on ``Sonoluminescence as Quantum Vacuum Radiation''}

We contest the recent claim that sonoluminescence may be explained in terms
of quantum vacuum radiation \cite{CE1}. Due to fundamental physical
limitations on bubble surface velocity, the predicted number of photons per
flash is indeed much smaller than unity. Therefore, quantum vacuum radiation
cannot be considered as an explanation of the observed sonoluminescence
phenomenon.

As the starting point of our critical comment we will use the theoretical
evaluation of radiation emitted into vacuum by a moving spherical interface
between two dielectric media \cite{CE1,CE2}.
In addition to the results presented
in these papers we will calculate the number of photons
emitted in each sonoluminescence cycle by
interpreting it in terms of parametric emission. Each
elementary radiation process corresponds to the emission of a pair of
photons $\left( \omega ,\omega ^{\prime }\right) $ carrying an energy $\hbar
(\omega +\omega ^{\prime })=\hbar \Omega $, where $\Omega $ is the frequency
of the mechanical motion. Equation (10) of \cite{CE1} or equation (4.8) of
\cite{CE2}, which give the energy radiated per pulsation of period $T$, are
thus translated into the following expression of the number of radiated
photons
\begin{equation}
{\cal N}=\alpha \int_0^\infty {\rm d}\Omega \ \Omega ^5\ \left| \int_0^T{\rm %
d}\tau \frac{R^2(\tau )}{c^2}e^{i\Omega \tau }\right| ^2  \label{NQR}
\end{equation}
$\alpha $ is a numerical factor ($\alpha \simeq 10^{-4}$ for a water-air
interface), and $R(\tau )$ is the time-dependent radius.

An estimate of ${\cal N}$ may be derived from a simple lorentzian model
(eq. (4.9) of \cite{CE2} with the same notations for $R_0$, $R_{\min }$ and
$\gamma$)
\begin{eqnarray}
&&{\cal N}=\frac{15\pi ^2}{16}\alpha \beta ^4  \nonumber \\
&&\beta ^2=\frac{R_0^2-R_{\min }^2}{c^2\gamma ^2}  \label{N}
\end{eqnarray}
The number of photons thus depends on a single parameter $\beta $ which is
the typical velocity of bubble surface measured with respect to the speed of
light. The time $\gamma $ characteristic of bubble collapse, which is
assumed to be much shorter than $T$, determines the frequency spectrum of
radiation but does not appear in the integrated photon number. A study of
the law of variation of velocity then shows that the photon number (\ref{N})
may not exceed a value determined by the maximal velocity
$\left( \frac{{\rm d}R}{{\rm d}t}\right) _{\max }$ of the bubble surface
\begin{equation}
{\cal N} \leq .1 \left( \frac{{\rm d}R}{c{\rm d}t}\right) _{\max }^4
\label{Nmax}
\end{equation}
{}For any maximal velocity smaller than the speed of light, the photon
number (\ref{N}) therefore remains smaller than unity. In contrast, observed
sonoluminescence corresponds to more than $10^5$ photons per flash (see for
instance Fig.3 of \cite{sono1}).

Moreover, the characteristic velocity of
the bubble surface corresponds to the sound velocity in air or water (see
Fig.4 of \cite{sono2}) rather than to the speed of light. For a maximal
velocity equal to sound velocity in water, ${\cal N}$ comes out as a very
small number of the order of $10^{-23}$ photons. We may emphasize that,
with a radius in the $\mu {\rm m}$ range (see Fig.3 of \cite{sono2}),
a velocity of the order
of $1{\rm km/s}$ leads to a typical time in the ${\rm ns}$ range, which fits
experimental reports (see again Fig.3 of \cite{sono2}). A much shorter time
scale
is used in \cite{CE1} in order to
explain the short duration of the flashes and the large width
of the thermal-like spectrum.
This very short time scale not only differs considerably
from the measured values, but it also corresponds to supraluminal velocities.

The precise values of the numerical factors which play a role
in the preceding discussion have been obtained from a specific model for the
variation
of the bubble radius. However, it can hardly be expected that a different
variation law of bubble radius may fill the gap between $10^{-23}$ and $10^5$
photons. The predicted number of photons per flash remains much smaller than
unity, which unavoidably leads to contest the claim that
``the theory of vacuum radiation seems to agree remarkably well
with the experimental results on sonoluminescence'' \cite{CE1}.

\vspace{1cm}
\noindent Astrid Lambrecht\\
Max-Planck-Institut f\"{u}r Quantenoptik\\
D-85748 Garching, Germany\\

\noindent Marc-Thierry Jaekel\\
Laboratoire de Physique Th\'{e}orique de l'Ecole Nor\-ma\-le
Su\-p\'{e}\-rieu\-re,
F-75231 Paris Cedex 05, France\\

\noindent Serge Reynaud\\
Laboratoire Kastler Brossel,
Universit\'{e} Pierre et Marie Curie,
F-75252 Paris Cedex 05, France\\

\noindent LPTENS 96/40


\end{document}